\documentclass[11pt,a4paper]{article}
\pdfoutput=1
\usepackage[english]{babel}
\usepackage{jheppub}

\makeatletter
\numberwithin{equation}{section}
\numberwithin{figure}{section}

\makeatother

\usepackage{babel}
\begin{document}

\title{General triple charged black ring solution in supergravity}
\author{Andrew Feldman}
\author{and Andrei A. Pomeransky}
\affiliation{Budker Institute of Nuclear Physics,\\
11, academician Lavrentiev Ave., Novosibirsk, Russia}
\affiliation{Physics Department, Novosibirsk State University,\\
2, Pirogova Str., Novosibirsk, Russia}
\emailAdd{andrew.l.feldman@gmail.com}
\emailAdd{pomeransky@gmail.com}
\abstract{
We present the general black ring solution in $U(1)^{3}$ supergravity in 5 dimensions with three independent dipole and electric charges. This immediately gives the general black ring solution in the minimal 5D supergravity as well.}
\maketitle

\section{Introduction}

\label{sec:intro}

Black rings are a type of black hole solutions in 5-dimensional
spacetime with the event horizon  homeomorphic to $S^{1}\times S^{2}\times R$ (see \cite{Emparan:2006} for a review). After the discovery of neutral black rings in pure gravity \cite{Emparan:2001wn} the
charged generalizations in supergravity were considered soon \cite{Elvang:2003yy,Elvang:2003mj,Emparan:2004wy,Elvang:2004xi}.
While the most general supersymmetric solution was found \cite{Elvang:2004rt,Bena:2004de,Elvang:2004ds,Gauntlett:2004qy},
the known families of nonextremal solutions lacked a number of free
parameters. In this paper we  present for the first time the general
nonextremal solution in $U(1)^{3}$ supergravity. The solution has
three independent electric and dipole charge parameters, plus 5 additional
independent parameters. The general solution has the usual Dirac-Misner
string and conical singularities. The condition of the absence of
singularities then singles out the family of regular solutions with
the complete set of independent parameters: 3 electric charges, 3
dipoles, 2 angular momenta and mass. As is well known, the corresponding solutions of the 5D minimal supergravity can be immediately obtained by setting the three electric charges and the three dipole charges equal, which makes all three gauge fields equal and all dilatons constant. 

Our starting point is the solution with a single nonzero gauge field
that was found in \cite{Chen:2012,Rocha:2012,Feldman:2012vd} using
the inverse scattering method \cite{Belinsky:1978,Belinsky:1979,Belinski:2001}.
It was pointed out in \cite{Rocha:2011} that when 2 gauge fields
and a dilaton of $U(1)^{3}$ supergravity are set equal to zero, the
result is the 5D Kaluza-Klein theory: the pure 6-dimensional gravity
compactified on a circle. And therefore, it was possible to apply
to this 6D pure gravity theory the procedure outlined in \cite{Pomeransky:2005}
based on the inverse scattering method and used in \cite{Pomeransky:2006}
to derive the general black ring solution in the 5D pure gravity.
In the concluding section of \cite{Feldman:2012vd} we suggested that the
missing parameters can be introduced into the solution by symmetrizing
it with respect to the dipole charges. The idea is that the general
solution should be invariant under the simultaneous permutations of
the gauge fields, dilatons and corresponding dipole charges. The observed
lack of symmetry is the consequence of 2 dipole charges parameters
being set to some particular value and when one eliminates this asymmetry
one gets the general solution. This turned out to be indeed possible
to do, but first one needs to eliminate another source of asymmetry
coming from a somewhat arbitrary choice of coordinates used in \cite{Feldman:2012vd}. We explain the details of this derivation in section \ref{sec:Der} after introducing various necessary notations in section \ref{sec:Not}. The solution is presented in section \ref{sec:Sol} using  a set of functions introduced in the Appendix \ref{sec:Set}. As we have checked numerically with the precision better than $10^{-100}$ the solution indeed satisfies the field equations written down in the Appendix \ref{sec:Equ}. In  the concluding section \ref{sec:Con} we discuss some possible directions for the future work.

\section{Notations}
\label{sec:Not}
Let us start by introducing the necessary notations for coordinates, parameters, fields and so on. The metric components,
as well as all other fields, depend only on two coordinates: $u$
and $v$. There are also three other coordinates in the 5D space-time:
$t,$ $\phi$, $\psi$ of which the metric is independent.
They correspond therefore to three commuting symmetries that can be
described by three commuting Killing vectors. There is also a not so
short list of parameters. First of all, the four parameters $x_{i}$,
$i=0,1,2,3$ give positions of poles in the inverse metric. These
poles are related to the very useful notion of rods \cite{Harmark:2004}.
By a M\"obius transformation of coordinates $u$ and $v$ one can take
three of the $x_{i}$ to the arbitrary values, while only the remaining
fourth constant is indeed a parameter. A popular choice is $x_{1}=-1$,
$x_{2}=1$, $x_{3}=-1/c$, $x_{0}=\infty$. Then there are three parameters
$a_{i},$ $i=1,2,3$ that are needed to obtain by imposing the necessary
regularity conditions the general nonsingular doubly rotating solution.
Furthermore, there are four parameters $y_{i}$, $i=0,1,2,3$. Three
of them are related to dipole charges, but $y_{0}$ is already present
in the Emparan-Reall neutral black ring with a single rotation. Finally,
one can apply a sequence of three boosts, interspersed with duality
transformations, to charge the solution with respect to three independent
electric charges. It is convenient to characterize the boosts by their
velocities $\beta_{i}$, $i=1,2,3$. 

Despite the large number of parameters it is possible to present the
solution in a relatively compact and readable form thanks to its numerous
symmetries. Let us consider a set of three transformations ($i=1,2,3$):
\begin{equation}
u\rightarrow h_{i}(v),\; v\rightarrow h_{i}(u),\; a_{i}\rightarrow1/a_{i}.\label{transform}
\end{equation}
Here $h_{1}(u)$ is the M\"obius transformation with the properties:
$h_{1}(x_{1})=x_{0}$, $h_{1}(x_{2})=x_{3}$ and $h_{1}(h_{1}(u))=u$.
The explicit expression is
\begin{equation}
h_{1}(u)=\frac{(x_{2}x_{3}-x_{0}x_{1})u+x_{0}x_{1}(x_{2}+x_{3})-x_{2}x_{3}(x_{0}+x_{1})}{(x_{2}+x_{3}-x_{0}-x_{1})u+x_{0}x_{1}-x_{2}x_{3}}\,.
\end{equation}
$h_{2}(u)$ and $h_{3}(u)$ can be obtained from $h_{1}(u)$ by exchanging
$x_{1}$ with $x_{2}$ and with $x_{3}$, respectively. All components
of all fields (metric, gauge fields and scalars) of the solution
do not change under the transformations \ref{transform}. In turn,
this invariance follows from the fact that all these components have
the form of ratios, and under the discussed transformation numerators
and denominators of the ratios are multiplied by the same common factor. The numerators and denominators are polynomials
in each $a_{i}$ of degree at most 2. The invariance under \ref{transform}
allows one to express the coefficient of $a_{i}^{2}$ in these polynomials in terms of the term of zero degree in $a_{i}$. To this end we introduce the
symmetrization operators $S_{i}$, which act in the space of functions
of $u$ and $v$ as follows:
\begin{equation}
S_{i}\{f(u,v)\}=f(u,v)-a_{i}^{2}f(h_{i}(v),h_{i}(u)).
\end{equation}
The operators $S_{i}$ commute with each other. The composition of
all $S_{i}$ will be denoted by $S\{f\}=S_{1}\{S_{2}\{S_{3}\{f\}\}\}$.

We never use summation over repeating indices, except when the sum
is written explicitly. We use the convention that the indices
$i$, $j$, $k$, $l$ have arbitrary but different values, in other
words they represent a permutation of $(0,1,2,3)$. We will use
frequently 6-component quantities, where each component corresponds
to an unordered pair of indices $i$ and $j$ ($i\ne j$). The addition
and multiplication for them is the component-wise one. The double
vertical line brackets $||...||$ will denote the sum of all 6 components
of the quantity in the brackets. We introduce two functions of a single
variable $l(z)$ and $q(z)$ that take such 6-component values: 
\begin{equation}
l_{ij}(z)=(z-x_{i})(z-x_{j}),\; q_{ij}(z)=\sqrt{z-x_{i}}\sqrt{z-x_{j}},
\end{equation}
and a 6-component valued function of two variables $r(u,v)$:
\begin{equation}
r_{ij}(u,v)=\frac{(u-x_{i})(v-x_{i})}{G^{\prime}(x_{i})}+\frac{(u-x_{j})(v-x_{j})}{G^{\prime}(x_{j})},
\end{equation}
where 
\begin{equation}
G(u)=\prod_{i=0}^{3}(u-x_{i}).
\end{equation}

Let us introduce also a constant 6-component quantity $\Delta$ with
the components equal to $\Delta_{ij}=(x_{i}-x_{j})^{2}$. For a 6-component
quantity we will denote by bar the transposition, which consists in
exchanging $ij$ with $kl$ components, for example: $\bar{q}_{ij}=q_{kl}$.
Note that $r_{ij}(u,v)=-r_{kl}(u,v)$, or with the above notation: $\bar{r}(u,v)=-r(u,v)$.
It is also useful to combine the constants $a_{i}$ into the 6-component
constant $a$ with the components $a_{0i}=a_{i},$ and the property
$\bar{a}=-a$. Let us introduce also a trilinear function of 3 variables
$c(p,s,t)$, where each of the variables $s$, $p$ and $t$ is a
6-component quantity, and the function is equal to
\begin{equation}
c(p,s,t)=\frac{1}{2}\sum_{i\ne j\ne k}p_{ij}s_{jk}t_{ki},\label{c}
\end{equation}
where the sum is over all 24 ordered triplets $(i,j,k)$ of non-equal
values of indices. Such function corresponds to a totally symmetric
tensor of rank 3. When one takes all three arguments equal, one obtains
a cubic function that we will call $c_{3}$: $c_{3}(t)=\frac{1}{3}c(t,t,t).$
We will use also a symmetrization operator $S^{\prime}$, which acts
on the 6-component quantities as follows:
\begin{equation}
S^{\prime}(t)_{0i} = S_{j}(S_{k}(t_{0i})),\;  S^{\prime}(t)_{ij}=S_{i}(S_{j}(t_{ij})),
\end{equation}
where the triple $\{i,j,k\}$ is a permutation of $\{1,2,3\}$.

Having introduced all these notations, we are now ready to write down the complete set of functions of coordinates and parameters which appear in the solution. This is done in the Appendix \ref{sec:Set}. This set is comprised of ten functions $H_{ij}$, $K_{ij}$, $\Omega^{ij},$ $\omega_{i}$,
$\Sigma_{i}$, $\Pi_{ij}$, $\Xi_{ij}$, $Q$, $Z$, and $g^{\varphi\psi}$.

\section{Derivation}

\label{sec:Der}
Let us now explain how we were able to recover the general family of solutions which is the subject of this paper, starting from its particular subfamily presented in our previous paper \cite{Feldman:2012vd}.  The solution in \cite{Feldman:2012vd} depends among others on 2 parameters $y_1$ and $y_2$. It will be convenient to shift the index and rename the parameters $y_0$ and $y_1$. Recall also, that in  \cite{Feldman:2012vd} we had the following choice of parameters that specify the positions of poles in the inverse metric: $x_i$ for $i=1,2,3$ in arbitrary positions, and the fourth pole fixed at the infinity: $x_0=\infty$. If one wants to put the solution in the maximally symmetric form, one would like to treat all $x_i$ on equal footing and make $x_0$ a free parameter too. This can be done by a coordinate transformation,  making substitutions $u\rightarrow h(u)$ and $v\rightarrow h(v)$, where $h(v)$ is a M\"obius transformation that has a pole at the point $x_0$: 
\begin{equation}
h(u)=\frac{\alpha u +\beta }{u-x_0}.
\end{equation}
One has to make also the same substitution for the constants: $x_i\rightarrow h(x_i)$, $i=1,2,3$ and $y_i\rightarrow h(y_i)$. At the same time one rescales the constants 
\begin{equation}
a_i\rightarrow a_i \frac{\beta + \alpha x_0}{x_i-x_0}\frac{\sqrt{x_i-y_0}\sqrt{x_i-y_1}}{\sqrt{x_0-y_0}\sqrt{x_0-y_1}}, 
\end{equation}
$i=1,2,3$ in order to eliminate everywhere the dependence on $\alpha$ and $\beta$ and to put $x_0$ on equal footing with other $x_i$.

In the new coordinates the solution has the following instructive property. Let us take $g^{\varphi\varphi}$ component as the simplest example. It contains a term with the following product $(u-x_0)^2(u-y_0)(u-y_1)$. One can say that the explicit dependence on $x_0$, which violates the symmetry among  all $x_i$, is the effect of setting $y_2=y_3=x_0$ in the symmetric general solution. We see that if one replaces $(u-x_0)^2$ by $(u-y_2)(u-y_3)$, one restores both symmetry and generality (hopefully) at the same time. We have applied this idea systematically to all components of inverse metric, gauge fields and scalars. A minor complication is that not all components should be symmetric in all $y_i$. Only $g^{\varphi\varphi}$, $g^{\psi\psi}$ and $g^{\varphi\psi}$ have this total symmetry. Other components of the inverse metric, namely  $g^{tt}$, $g^{t\varphi}$ and $g^{t\psi}$, single out $y_0$ but should be symmetric in $y_1$, $y_2$ and $y_3$. The gauge fields $A_j$ single out the corresponding $y_j$ and $y_0$, but should be symmetric in the remaining two $y_i$ ($i\neq j\neq 0$). The same is true for scalar fields. After one gains some experience, the symmetrization procedure becomes almost straightforward, only with a small amount of guesswork needed. Fortunately, there is an excellent way to check the correctness for each component of the inverse metric separately: the residues at the poles in  $u$ should not depend on $v$ apart from a common factor which is easy to cancel completely. Furthermore, the residues should factorize: $res(g^{ij})\sim \rho^i\rho^j$, where $\rho^i$ is a constant vector -- the rod direction (for the description of rods see \cite{Harmark:2004}). This test is very strict because it is extremely improbable to have such factorization to hold by chance in an incorrect expression. When all components of the inverse metric have been obtained, one more test becomes available: it turns out that the determinant of the $3 \times 3$ matrix of inverse metric components for coordinates $t$, $\varphi$ and $\psi$ has a very simple form $\det(g^{ij})=\frac{(u-v)^{4}}{G(u)G(v)}$.
After all fields have been found, one can finally check that they indeed satisfy the field equations. Due to the high enough complexity of the solution,  we were not able to perform this check analytically even with the help of a computer algebra system. Instead, it was possible to do this numerically with a precision of more than 100 digits. Such numerical precision  is absolutely sufficient to convince everyone that the solution is correct. We used Wolfram Mathematica for both algebraic manipulations and numerical calculations. A Mathematica notebook that contains the solution and the numerical checks of the field equations is available on request.

Once the solution with general values of the dipole charges is found one can turn on the electric charges too. There is a well-known procedure for charging a 5D $U(1)^3$ supergravity solution (see e.g. \cite{Elvang:2004xi,Hoskisson:2008qq,Galtsov:2009,Breckenridge:1996is}).
It can be done by uplifting the solution to six dimensions, treating one of the gauge fields as a Kaluza-Klein one, and making a boost along the compact sixth direction. The other two gauge fields at the same time combine into the 2-from. Then one can reduce the result back to five dimensions and repeat the procedure with the next gauge field. After three boosts one obtains the general solution with three independent electric charges. The charges are parametrized by the velocities $\beta_i$ of the boosts.

\section{Solution}
\label{sec:Sol}
In this section we will present the general black ring solution of the 5D $U(1)^3$ supergravity field equations. The field equations themselves are  written down in the Appendix \ref{sec:Equ}. We will express the components of the fields in terms of a set of auxiliary functions defined in the Appendix \ref{sec:Set}. Let us start from the scalar fields $\Phi_i$. They have the following form:
\begin{equation}
e^{\Phi_i}=\frac{\chi_i}{\chi},\; \chi_{i}=H_{0i}-\beta_{i}^{2}\tilde{H}_{0i}+2\beta_{i}K_{0i},\;\chi=\prod_{i=1}^{3}\chi_{i}^{1/3}\,.
\end{equation}
The $tt$ component of the inverse metric is:
\[
g^{tt}=\chi^{-1}\left(\sum_{m,n=0}^{3}\left(-2\beta_{0}\frac{\beta_{n}}{\beta_{m}}\Pi_{mn}-\beta_{0}^{2}\frac{\Xi_{mn}}{\beta_{m}\beta_{n}}\right)+\sum_{m=0}^{3}\left(\beta_{0}^{2}\frac{\Sigma_{m}}{\beta_{m}^{2}}-\beta_{m}^{2}\tilde{\Sigma}_{m}\right)+2\beta_{0}Q\right),
\]
where the diagonal elements of $\Pi$ and $\Xi$ are defined to be zero: $\Pi_{mm}=\Xi_{mm}=0$ and a shorthand notation $\beta_{0}=-\beta_{1}\beta_{2}\beta_{3}$ is introduced. The other non-zero inverse metric components are:
\begin{equation}
g^{t\varphi}=\chi^{-1}\sum_{m=0}^{3}\left(\beta_{m}\tilde{\omega}_{m}^{\psi}+\frac{\beta_{0}}{\beta_{m}}\omega_{m}^{\varphi}\right),\; g^{\varphi\varphi}=\frac{Z}{\chi},\; g^{\psi\psi}=-\frac{\tilde{Z}}{\chi},
\end{equation}
\begin{eqnarray}
g^{\varphi\psi} & = & -\chi^{-1}||a\,\frac{r(u,v)}{u-v}\, q(y_{0})q(y_{1})q(y_{2})q(y_{3})S^{\prime}\left\{ \frac{(u-v)^{2}}{l(u)l(v)}\right\} ||\nonumber \\
 & + & \frac{u-v}{\chi G(u)G(v)}\,\prod_{m=0}^{3}G(y_{m})\, c_{3}\left(\frac{a\, r(u,v)\Delta}{q(y_{0})q(y_{1})q(y_{2})q(y_{3})}\right).
\end{eqnarray}
\begin{equation}
g^{uu}=\frac{(u-v)^2}{C_0\chi}G(u),\;g^{vv}=-\frac{(u-v)^2}{C_0\chi}G(v),
\end{equation}
where $C_0$ is an arbitrary constant.

The determinant of the $3\times 3$ matrix of the inverse metric components $g^{mn}$ ( $m,n=t,\varphi,\psi$) has the simple form:
\begin{equation}
\det(g^{mn})=\frac{(u-v)^{4}}{G(u)G(v)}.
\end{equation}

The gauge vector potentials are:
\begin{eqnarray*}
A_{t}^{i} & =  &\frac{1}{\chi_{i}} \left(\frac{1+\beta_{i}^{2}}{1-\beta_{i}^{2}}K_{0i}+\frac{\beta_{i}}{1-\beta_{i}^{2}}\left(H_{0i}-\tilde{H}_{0i}\right)\right),\\
A_{\varphi}^{i} & = & \frac{1}{\chi_{i}}\left(\Omega_{\varphi}^{i0}-\beta_{j}\beta_{k}\Omega_{\varphi}^{0i}+\beta_{i}\beta_{j}\Omega_{\varphi}^{jk}+\beta_{i}\beta_{k}\Omega_{\varphi}^{kj}+\beta_{1}\beta_{2}\beta_{3}\tilde{\Omega}_{\psi}^{i0}-\beta_{i}\tilde{\Omega}_{\psi}^{0i}+\beta_{k}\tilde{\Omega}_{\psi}^{jk}+\beta_{j}\tilde{\Omega}_{\psi}^{kj}\right).
\end{eqnarray*}
$A_{\psi}^{i}$ can be obtained from $A_{\varphi}^{i}$ by exchanging all $\phi$ and $\psi$ subscripts in the expression above.

\section{Conclusions}
\label{sec:Con}
In this paper we presented the general black ring solution in $U(1)^{3}$ supergravity (and therefore in the minimal 5D supergravity as well). We have tried to simplify it as much as possible. To this end we introduced several notations, which allowed to reduce the length of expressions considerably. Still we are not completely satisfied at this point with the form of the solution. One can hope that there is a formulation  that is both elegant and allows to check the validity of the solution analytically, instead of checking it numerically, as we were forced to do. Such formulation would uncover the natural algebraic structure of the solution and give it a nice mathematical sense. 

One possible way to reach such better understanding of the solution is to try to generalize it to a larger supergravity that reduces to the $U(1)^{3}$ theory when some fields vanish. Interesting examples are 11D supergravity (the low energy limit of M-theory) reduced to 5D on $T^6$ or on $K3 \times T^2$ which gives theories with 27 gauge fields \cite{Emparan:2008,Cvetic:1996xz}. In the first case some of the missing dipole charges can be generated by suitably uplifting our solution to 11D and then rotating it in the six compact directions. Combining this rotations with duality transformations can probably generate even more independent dipole charges. 

As an intermediate step one could also try to find the black ring solutions in the theory with just one additional gauge field considered in \cite{Breckenridge:1996is}. It would be straightforward to add an electric charge with respect to the additional gauge field. Adding the dipole charge using symmetry considerations is not straightforward, but may turn out to be possible with some luck. Another interesting problem is a deeper investigation of the black ring solution in minimal 5D supergravity. In this case the number of independent parameters and fields is smaller than in more general case of $U(1)^3$ supergravity, and therefore one can hope to be able to get more explicit expressions for the regular solution and its  mass, angular momenta etc.

\begin{acknowledgments}
A.F. acknowledges the financial support by the Dynasty Foundation. 
\end{acknowledgments}

\providecommand{\href}[2]{#2}\begingroup\raggedright\endgroup

\appendix

\section{Field equations}
\label{sec:Equ}
In this appendix the Greek indices enumerate the five-dimensional spacetime coordinates, and the summation over repeating Greek indices is assumed.
The field equations  for the $U(1)^3$ five-dimensional supergravity can be derived from the following action:
\begin{equation}\label{action}
I=\int d^5 x \sqrt{-g}\left( R -\frac{1}{4}\sum_{i=1}^{3}e^{2 \Phi_i} F^i_{\mu\nu}  F^{i\mu\nu}\,-\frac{1}{2}\sum_{i=1}^{3} g^{\mu\nu}\partial_{\mu}\Phi_i \partial_{\nu}\Phi_i\right) -\int dA^1 \wedge dA^2 \wedge A^3,
\end{equation}
with the constraint $\Phi_1+\Phi_2+\Phi_3=0$ and where $F^i=dA^i$.
The resulting field equations have the form
\begin{eqnarray}
\partial_\nu \left(\sqrt{-g} e^{a \Phi_i} F^{i\sigma\nu}\right) &=& \frac{1}{4} \epsilon^{\mu\nu\kappa\lambda\sigma} F^j_{\mu\nu} F^k_{\kappa\lambda},\nonumber\\
\partial_\mu \left( \sqrt{-g} g^{\mu\nu}\partial_\nu \Phi_i \right) &=& \frac{\sqrt{-g}}{6}  \left(2e^{2\Phi_i} F^i_{\mu\nu}  F^{i\mu\nu}-e^{2\Phi_j} F^j_{\mu\nu}  F^{j\mu\nu}-e^{2\Phi_k} F^k_{\mu\nu}  F^{k\mu\nu}\right),\nonumber\\
R_{\mu\nu}-\frac{1}{2}g_{\mu\nu}R &=&\frac{1}{2} \sum_{i=1}^{3}e^{2 \Phi_i}\left(F^{i}_{\lambda\mu}F^{i\lambda}{}_{\nu}
-\frac{1}{4}g_{\mu\nu}F^i_{\kappa\lambda} F^{i\kappa\lambda}\right)\\
&+& \frac{1}{2}\sum_{i=1}^{3}\left( \partial_{\mu}\Phi_i \partial_{\nu}\Phi_i-\frac{1}{2}g_{\mu\nu} g^{\kappa\lambda}\partial_{\kappa}\Phi_i \partial_{\lambda}\Phi_i \right).\nonumber
\end{eqnarray}
Let us write down the equations for the gauge field strength more explicitly:
\begin{eqnarray}
& & \partial_u\left(\frac{G(u)}{(u-v)^2} e^{2 \Phi_i}\sum_{n=t,\varphi,\psi}g^{mn}\partial_u A^i_n-  \sum_{n=t,\varphi,\psi}\sum_{s=t,\varphi,\psi}\epsilon^{mns} A^j_n \partial_v A^k_s\right)\nonumber\\
&=& \partial_v\left(\frac{G(v)}{(u-v)^2} e^{2 \Phi_i}\sum_{n=t,\varphi,\psi}g^{mn}\partial_v A^i_n-\sum_{n=t,\varphi,\psi}\sum_{s=t,\varphi,\psi}\epsilon^{mns} A^j_n \partial_u A^k_s\right)\,,
\end{eqnarray}
where $\epsilon^{mns}$ is the antisymmetric tensor and $\epsilon^{t\varphi\psi}=1$. The field equations for the dilatons can be reduced to the statement, that the expression
\begin{eqnarray}
& &\partial_u \left(\frac{G(u)}{(u-v)^2}\partial_u \Phi_i \right)-
\partial_v \left(\frac{G(v)}{(u-v)^2}\partial_v \Phi_i \right)\nonumber\\
&-&\sum_{m,n=t,\varphi,\psi}e^{2\Phi_i}\left(\frac{G(u)}{(u-v)^2}g^{mn}\partial_u A^i_m \partial_u A^i_n-
\frac{G(v)}{(u-v)^2}g^{mn}\partial_v A^i_m \partial_v A^i_n\right)
\end{eqnarray}
does not depend on $i$ and the constraint $\Phi_1+\Phi_2+\Phi_3=0$ is satisfied. Finally, the Einstein equations can be reduced to the following form:
\begin{eqnarray}
& & \partial_u\left(\frac{G(u)}{(u-v)^2}\sum_{s=t,\varphi,\psi}g_{ms}\partial_u g^{sn}\right) -\partial_v\left(\frac{G(v)}{(u-v)^2}\sum_{s=t,\varphi,\psi}g_{ms}\partial_v g^{sn}\right)\nonumber\\
&=&\sum_{i=1}^{3}\sum_{s=t,\varphi,\psi}e^{2 \Phi_i}\left( \frac{G(u)}{(u-v)^2}\partial_u A^i_m g^{ns}\partial_u A^i_s-\frac{G(v)}{(u-v)^2}\partial_v A^i_m g^{ns}\partial_v A^i_s\right)\nonumber\\
&-&\frac{1}{3}\delta^n_m\sum_{i=1}^{3}\sum_{p,s=t,\varphi,\psi}e^{2 \Phi_i}\left( \frac{G(u)}{(u-v)^2}\partial_u A^i_p g^{ps}\partial_u A^i_s-\frac{G(v)}{(u-v)^2}\partial_v A^i_p g^{ps}\partial_v A^i_s\right)
\end{eqnarray}

\section{Set of Functions}
\label{sec:Set}
In this appendix we define the set of function used in section \ref{sec:Sol} to write down the black ring solution. The notations used here were described in section \ref{sec:Not}.
\begin{eqnarray}
H_{ij} & = & -S\left\{ \frac{1}{(u-v)^{2}}(u-y_{i})(u-y_{j})(v-y_{k})(v-y_{l})\right\} \nonumber \\
 & + & \frac{1}{(u-v)^{2}}c\left(a\, r(u,v),a\, r(u,v),\Delta q(y_{i})q(y_{j})q(y_{k})q(y_{l})\right).\label{H}
\end{eqnarray}
\begin{equation}
K_{ij}=||S^{\prime}\left\{ \frac{a\, r(u,v)}{u-v}\right\} q(y_{i})q(y_{j})\bar{q}(y_{k})\bar{q}(y_{l})||.\label{K}
\end{equation}
\begin{eqnarray}
\Omega_{\varphi}^{ij} & = & ||\frac{a\, r(u,v)}{u-v}\, q(y_{i})\bar{q}(y_{j})\bar{q}(y_{k})\bar{q}(y_{l})S^{\prime}\left\{ u-y_{j}-\frac{l(u)}{u-v}\right\} ||,\nonumber \\
\Omega_{\psi}^{ij} & = & -\prod_{m=1}^{4}\sqrt{x_{m}-y_{i}}\frac{\partial H_{ij}}{\partial y_{i}}.
\end{eqnarray}
\begin{eqnarray}
Z & = & -S\left\{ \frac{1}{G(v)}(v-y_{0})(v-y_{1})(v-y_{2})(v-y_{3})\right\} \nonumber \\
 & + & c\left(a\, r(u,v),a\, r(u,v),\bar{\Delta}q(y_{0})q(y_{1})q(y_{2})q(y_{3})\left(\frac{1}{\bar{l}(u)l(v)}-\frac{a^{2}}{l(u)\bar{l}(v)}\right)\right)\,.
\end{eqnarray}
\begin{eqnarray}
\omega_{i}^{\varphi} & = & \prod_{m=1}^{4}\sqrt{x_{m}-y_{i}}\frac{\partial Z}{\partial y_{i}},\nonumber \\
\omega_{i}^{\psi} & = & ||\frac{a\, r(u,v)}{u-v}\,\bar{q}(y_{i})q(y_{j})q(y_{k})q(y_{l})S^{\prime}\left\{ \frac{u-v}{l(v)}(1-\frac{u-v}{l(u)}(u-y_{i}))\right\} ||\nonumber \\
 & - & \frac{u-v}{G(u)G(v)}\,\prod_{m=1}^{4}\sqrt{x_{m}-y_{i}}\, G(y_{j})G(y_{k})G(y_{l})c_{3}(\frac{a\, r(u,v)\Delta}{q(y_{j})q(y_{k})q(y_{l})}).
\end{eqnarray}
 \begin{eqnarray*}
\Sigma_{i} & = & S\left\{ \left(\frac{(u-y_{i})^{2}}{(u-v)^{2}}-\frac{G(y_{i})}{G(v)}\right)\frac{(v-y_{j})(v-y_{k})(v-y_{l})}{(v-y_{i})}\right\} \\
 & - & c\left(a\, r(u,v),a\, r(u,v),\Delta q(y_{1})q(y_{2})q(y_{3})q(y_{4})\left(\frac{1-a^{2}}{(u-v)^{2}}-\bar{l}(y_{i})\left(\frac{1}{\bar{l}(u)l(v)}-\frac{a^{2}}{l(u)\bar{l}(v)}\right)\right)\right).
\end{eqnarray*}
\begin{eqnarray*}
\Pi_{ij} & = & ||\frac{a\, r(u,v)}{u-v}\,\bar{q}(y_{i})\bar{q}(y_{j})q(y_{k})q(y_{l})S^{\prime}\left\{ \left(1-\frac{u-v}{l(u)}(u-y_{i})\right)\left(1-\frac{v-u}{l(v)}(v-y_{j})\right)\right\} ||\\
 & + & \frac{u-v}{G(u)G(v)}\, G(y_{k})G(y_{l})\prod_{m=1}^{4}\sqrt{x_{m}-y_{i}}\sqrt{x_{m}-y_{j}}\, c_{3}\left(\frac{a\, r(u,v)\Delta}{q(y_{k})q(y_{l})}\right).
\end{eqnarray*}
\begin{eqnarray}
\Xi_{ij} & = & -\prod_{m=1}^{4}\sqrt{x_{m}-y_{i}}\sqrt{x_{m}-y_{j}}\frac{\partial^{2}Z}{\partial y_{i}\partial y_{j}}-\prod_{m=1}^{4}\sqrt{x_{m}-y_{k}}\sqrt{x_{m}-y_{l}}\frac{\partial^{2}Z}{\partial y_{k}\partial y_{l}}\nonumber \\
 & + & \frac{1}{2}c\left(a\, r(u,v),a\, r(u,v),\Delta\,\bar{\Delta}\, q(y_{i})q(y_{j})\bar{q}(y_{k})\bar{q}(y_{l})\left(\frac{1}{\bar{l}(u)l(v)}-\frac{a^{2}}{l(u)\bar{l}(v)}\right)\right).
\end{eqnarray}
\begin{eqnarray*}
Q & = & ||a\,\frac{r(u,v)}{u-v}q(y_{0})q(y_{1})q(y_{2})q(y_{3})(\bar{l}(y_{0})+\bar{l}(y_{1})+\bar{l}(y_{2})+\bar{l}(y_{3}))S^{\prime}\left\{ \frac{(u-v)^{2}}{l(u)l(v)}\right\} ||\\
 & - & \frac{u-v}{G(u)G(v)}\,\prod_{m=0}^{3}G(y_{m})\sum_{n=0}^{3}\frac{1}{G(y_{n})}c_{3}\left(\frac{a\, r(u,v)q(y_{n})\Delta}{q(y_{0})q(y_{1})q(y_{2})q(y_{3})}\right).
\end{eqnarray*}
\end{document}